\documentclass[prd,aps,letterpaper,floatfix,superscriptaddress,preprintnumbers,11pt,nofootinbib]{revtex4}
\usepackage{amsmath}
\usepackage{mathrsfs}
\usepackage{graphicx}  
\usepackage{dcolumn}   
\usepackage{bm}        
\usepackage{amssymb}   
\usepackage{amsmath}
\usepackage{multirow}
\usepackage{slashed}
\usepackage{braket, slashed, bm}
\usepackage{array,multirow}
\usepackage[normalem]{ulem}
\usepackage{xcolor, cancel,youngtab}
\usepackage{comment}
\usepackage[inkscapeformat=png]{svg}

\usepackage[T1]{fontenc} 

\usepackage{mathrsfs}

\usepackage{booktabs}
\usepackage{adjustbox}
\usepackage{mathtools}

\usepackage{soul}

\usepackage{tikz}

\definecolor{labelkey}{rgb}{0,0.5,0.0}

\usepackage{xparse}
\usepackage{etoolbox}

\usepackage{feynmp}
\usepackage{graphicx,psfrag,amsmath}
\usepackage{color}
\usepackage{slashed}
\usepackage{xcolor}
\usepackage{multirow}
\usepackage{ulem}
\usepackage{xcolor}
\usepackage{natbib}
\usepackage{notoccite}
\usepackage[colorlinks]{hyperref}
\hypersetup{ 
    colorlinks=true,
    linkcolor=blue,
    filecolor=magenta,      
    urlcolor=cyan,
}
\usepackage{hypcap}
\def\lsim{\mathrel{\raise.3ex\hbox{$<$\kern-.75em\lower1ex\hbox{$\sim$}}}}
\def\gsim{\mathrel{\raise.3ex\hbox{$>$\kern-.75em\lower1ex\hbox{$\sim$}}}}

\evensidemargin=\oddsidemargin
\newcommand{\be}{\begin{equation}} 
\newcommand{\ee}{\end{equation}} 
\newcommand{\bea}{\begin{eqnarray}}
\newcommand{\eea}{\end{eqnarray}}
\newcommand{\bee}{\begin{eqnarray*}}
	\newcommand{\eee}{\end{eqnarray*}}

\renewcommand{\O}{\mathcal{O}}

\NewDocumentCommand{\Op}{ m m O{} o }{
	\O^{\ifblank{#3}{}{#3,}#2 }_{\IfNoValueTF{#4}{#1}{\substack{#1\\#4}}}
}
\NewDocumentCommand{\lwc}{ m m O{} o }{
	L^{\ifblank{#3}{}{#3,}#2 }_{\IfNoValueTF{#4}{#1}{\substack{#1\\#4}}}
}
\NewDocumentCommand{\dlwc}{ m m O{} o }{
	{\dot L}^{\ifblank{#3}{}{#3,}#2 }_{\IfNoValueTF{#4}{#1}{\substack{#1\\#4}}}
}

\begin{document}
\title{Constraining anti-baryonic dark matter through correlated nucleon decay signatures}
  \author{Mathew Thomas Arun}
\email{mathewthomas@iisertvm.ac.in}
\affiliation{School of Physics, Indian Institute of Science Education and Research, Thiruvananthapuram, Kerala 695551, India}
\author{Anuja Bandu Khadse}
\affiliation{Indian Institute of Science Education and Research, Pune, Maharashtra 411008, India}

\begin{abstract}
Baryon number violation in the visible sector induced by anti-baryonic dark matter provides a viable mechanism for low-scale baryogenesis. Two of the most sensitive probes of this scenario are neutron decay processes such as $n \to \bar{\nu} + \text{invisible}$ and $n \to \pi^0 + \text{invisible}$. In this work, we discuss the possible spontaneous breaking of baryon symmetry in the dark sector and the generation of di-nucleon decay processes such as $nn \to \bar{\nu}\bar{\nu}$ and $nn \to \pi^0\pi^0$ at one-loop, arising from the operators responsible for induced nucleon decays. While the induced nucleon decay rates in this model depend on the dark matter density, di-nucleon decay processes do not, providing a complementary probe of the new physics. We thus use nucleon and di-nucleon decay bounds to constrain the local density and mass of the anti-baryonic dark matter.
\end{abstract}

\maketitle

\section{Introduction}
Baryon asymmetry and dark matter remain two of the most profound mysteries of our universe. While extensive cosmological evidence, such as the cosmic microwave background anisotropies and large-scale structure formation~\cite{Planck:2019nip, Migliaccio:2019gbz}, supports their existence, the underlying new physics continues to evade terrestrial experiments like XENONnT~\cite{aprile2023first}, LUX-ZEPLIN (LZ)~\cite{aalbers2023background}, KamLAND~\cite{KamLAND:2005pen}, Super-Kamiokande~\cite{Super-Kamiokande:2013rwg, abe2015search, abe2017search,takeuchi2020recent}, JUNO~\cite{JUNO:2024pur} and indirect searches for particle fluxes from galaxies~\cite{Fermi-LAT:2020pst}. Moreover, the observed baryon asymmetry is difficult to reconcile with Standard Model (SM) interactions, as electroweak baryogenesis requires additional CP violation and baryon number violation beyond the SM~\cite{morrissey2012electroweak}. 

In recent years, low-scale dark matter-induced baryon number violation~\cite{Kaplan:2009ag,Haba:2010bm,Davoudiasl:2010am,Davoudiasl:2011fj,petraki2013review,Huang:2013xfa,Arun:2022eqs,Akshay:2022vkb} has emerged as a compelling framework to bridge this gap against the backdrop of baryogenesis mechanisms involving high-scale interactions~\cite{Davidson:2008bu,Grossman:2018rdg}. Such frameworks often predict observable induced nucleon decay channels with new physics interactions of dark matter charged with anti-baryon number that allow for baryon number violation at experimentally accessible scales, unlike the severely restricted baryon number violation at  Grand Unification scales. In these "asymmetric" dark matter models, proton and neutron decay can be induced typically via non-renormalizable operators like,
\begin{eqnarray}
    \mathcal{O}_{uude^+} = \frac{1}{\Lambda_{uud \rightarrow e^+}^4} \hspace{0.1cm} u_r d_r Q_l L_l \Phi_i^{\dagger} \Phi_j  \ , &\qquad&
    \mathcal{O}_{uud \pi^+} = \frac{1}{\Lambda_{uud \rightarrow \pi^+}^3} \hspace{0.1cm} u_r u_r d_r  \Psi_i \Phi_j^{\dagger} \ , \nonumber \\
    \mathcal{O}_{udd\nu} = \frac{1}{\Lambda_{udd \rightarrow \nu}^4} \hspace{0.1cm} u_r d_r Q_l L_l \Phi_i^{\dagger} \Phi_j  \ , &\qquad&
    \mathcal{O}_{udd \pi^0} = \frac{1}{\Lambda_{udd \rightarrow \pi^0}^3} \hspace{0.1cm} u_r d_r d_r  \Psi_i \Phi_j^{\dagger} \ ,
\end{eqnarray}
where $\Phi_i$ and $\Psi_i$ are dark sector scalar and fermionic fields, and $Q_l, L_l$ are left-handed SM doublets, while $u_r, d_r$ are right-handed SM singlets. The dark sector fields are Standard Model gauge singlets, however, they are charged under the baryon number, $U(1)_{B}$, carrying anti-baryonic charge and/or under the Lepton number, $U(1)_{L}$, carrying anti-leptonic charge. In this work, the $U(1)_B$ and $U(1)_L$ groups are assumed to be gauged, and the associated anomalies are canceled either by a minimal expansion of the fermionic sector~\cite{FileviezPerez:2010gw,FileviezPerez:2024fzc} or by Green-Schwarz mechanism by including axion interactions~\cite{Green:1984sg,Anastasopoulos:2006cz,Coriano:2007fw}. Note that, above, we have considered only the minimal set of operators that preserve (violate) $B-L$ in the visible sector on the left (right). 

Since the experiments that we are interested in, which constrain these processes strongly, are at low energies, the quark level operators are matched to the dimension-5 hadronic operators~\cite{Claudson:1981gh},
\begin{eqnarray}
    \mathcal{O}_{pe^+} = \frac{1}{\Lambda_{p \rightarrow e^+}} \hspace{0.1cm} p e_l \Phi_1^{\dagger} \Phi_2 \ , &\qquad&
    \mathcal{O}_{p \pi^+} = \frac{1}{\Lambda_{p \rightarrow \pi^+}} \hspace{0.1cm} p \pi^+ \Psi \Phi^{\dagger} \ , \nonumber \\
    \mathcal{O}_{n \nu} = \frac{1}{\Lambda_{n \rightarrow \nu}} \hspace{0.1cm} n \nu_l \Phi_1^{\dagger} \Phi_2 \ , &\qquad& 
    \mathcal{O}_{n \pi^0} = \frac{1}{\Lambda_{n \rightarrow \pi^0}} \hspace{0.1cm} n \pi^0 \Psi \Phi^{\dagger} \ .
    \label{eq:hadronindBNV}
\end{eqnarray}
The stability of the particles involved in these processes are ensured if we assume the condition,
\begin{eqnarray}
|m_{\Phi_1}-m_{\Phi_2}| &<& m_p+m_e < m_{\Phi_1} + m_{\Phi_2} \nonumber \\ 
|m_{\Phi}-m_{\Psi}| &<& m_p+m_\pi < m_{\Phi} + m_{\Psi} \ ,
\label{eq:masscons}
\end{eqnarray}
where these are the masses of proton, electron, pion and the dark matter candidates. Similarly for neutrons also.
These operators generate `induced' decay channels like, 
\begin{eqnarray}
p + \Phi_1 \rightarrow e^+ + \Phi_2 \ , &\qquad& p + \Phi \rightarrow \pi^+ + \Psi \ , \nonumber \\
n + \Phi_1 \rightarrow \bar{\nu} + \Phi_2 \ , &\qquad& n + \Phi \rightarrow \pi^0 + \Psi \ .
\label{eq:DMprocesses}
\end{eqnarray}
On the other hand, next-order baryon number-violating (\(\Delta B=2\)) processes, such as neutron-antineutron oscillation~\cite{Super-Kamiokande:2020bov}, di-nucleon decay~\cite{Super-Kamiokande:2015jbb, Super-Kamiokande:2018apg}, exotic neutron disappearance modes~\cite{JUNO:2024pur}, and hydrogen–anti-hydrogen oscillation~\cite{Grossman:2018rdg}, are generated by higher mass dimension operators and are subject to constraints on the scale, typically, around 600 TeV or lower. The recent identification of 11 neutron-antineutron oscillation candidates against a background of \(9.3\pm2.7\) at Super-Kamiokande~\cite{Super-Kamiokande:2020bov} (0.37 megaton-year exposure) and upcoming searches at Hyper-Kamiokande~\cite{Hyper-Kamiokande:2018ofw}, DUNE~\cite{DUNE:2020lwj}, and HIBEAM/NNBAR ~\cite{Santoro:2022qvb} further motivate these studies.

Given that asymmetric dark matter models naturally couple the dark sector to baryon number violation in the visible sector, and that next-generation searches are sensitive to $\Delta B = 2$ processes, it is well motivated to ask whether dark sector induced baryon number violation in the visible sector correlates with $\Delta B = 2$ decays. Any confirmed signal of neutron-antineutron oscillations or of di-nucleon decay with back-to-back Cherenkov rings would point to spontaneous breaking of baryon number by two units in the dark sector~\cite{Arun:2022eqs}. In this framework, the same spurion of $\Delta B = 1$ breaking typically generates di-nucleon decays at one loop via the dimension-5 hadronic operators in Eq.~(\ref{eq:hadronindBNV}), thereby linking the $\Delta B = 2$ nuclear decay rates to the underlying symmetry-breaking parameters of the dark sector.


In this work, we provide a model for such breaking of baryon symmetry in the asymmetric dark sector and examine the implications of dark matter interactions on baryon number violation, focusing on the processes  $n\to \bar{\nu} + \text{invisible} $, $n \to \pi^0 + \text{invisible}$, $nn\to \bar{\nu}\bar{\nu}$ and $nn\to \pi^0\pi^0$. The experimental data relevant for this study are given in Table.~(\ref{tab:processes}). Using measurements of these nucleon and di-nucleon processes, we find that the mass and density of anti-baryonic dark matter at terrestrial experiments is constrained. 

\begin{table}[h]
    \centering
    \begin{tabular}{|c|c|c|}
    \hline 
   Process  & \hspace{0.2cm} Lifetime \hspace{0.2cm }& \hspace{0.2cm }Decay Width \hspace{0.2cm }\\ 
   & ($\mathsf{years}$) & ($\mathsf{GeV}$) \\ \hline 
    \hspace{0.2cm} $n\to \text{invisible} $\hspace{0.2cm} & \hspace{0.2cm} $5.8\times 10^{29}$~\cite{KamLAND:2005pen}\hspace{0.2cm} & $3.60 \times 10^{-62}$ \\[3pt]
    \hspace{0.2cm} $n \to \pi^0 + \text{invisible}$ \hspace{0.2cm} & $1.1\times 10^{33}$~\cite{Super-Kamiokande:2013rwg} & $1.90 \times 10^{-65}$ \\[3pt]
    \hspace{0.2cm} $nn \to \bar{\nu}\bar{\nu}$\hspace{0.2cm} & $1.4 \times 10^{30}$~\cite{KamLAND:2005pen} & $1.49 \times 10^{-62}$\\[3pt]
    \hspace{0.2cm} $nn \to \pi^0 \pi^0$\hspace{0.2cm} & $4.04 \times 10^{32}$~\cite{Super-Kamiokande:2015jbb} & $5.16 \times 10^{-63}$\\[3pt] \hline 
    \end{tabular}
    \caption{Limit on lifetimes and decay widths for the relevant processes from experiments.}
    \label{tab:processes}
\end{table}

The article is presented as follows. In section~\ref{sec:indbnv}, we discuss the dark sector interactions with visible sector baryon number violating operators, emphasizing on two different dark matter models that lead to $n \to \bar{\nu}$, which keeps the $B-L$ preserved, and $n\to \pi^0$, which violates $B$. In section~\ref{sec:SSB}, we develop models for spontaneous breaking of the dark sector baryon charge and derive the di-nucleon decay operators that are generated at one-loop from the induced baryon number violating operators. We derive, in section~\ref{sec:results}, the cross-sections for the induced baryon number violating processes, and correlate them with the matching provided at one-loop with the di-nucleon decay and compare them with the experimental values provided in Table.~(\ref{tab:processes}). We then rewrite these results as a function of dark matter density and mass of dark matter and provide limits that satisfy the di-nucleon processes and local density bound. In section~\ref{sec:summary} we summarize our findings.

\section{Dark matter induced nucleon decay}
\label{sec:indbnv}

In this section, we briefly recap the relevant concepts~\cite{Davoudiasl:2010am, Davoudiasl:2011fj, Huang:2013xfa} concerning induced baryon number violation by dark matter carrying anti-baryon number. Although both protons and neutrons can participate in these processes, we restrict our analysis to neutrons. The induced nucleon decay rate is given by,
\begin{equation}\label{decay}
    \Gamma_{IND} = \frac{\rho_{DM}}{M_{DM}} \hspace{0.1cm} (\sigma v)_{IND} \ ,
\end{equation}
where, $\rho_{DM}$ is the density of the dark matter, $M_{DM}$ is the mass of dark matter and $(\sigma v)$ is the thermal cross-section for the particular induced nucleon decay generated by the operators in Eq.~(\ref{eq:hadronindBNV}). In the following, we consider two scenarios and compute the corresponding induced thermal cross-section, $(\sigma v)_{IND}$.


\subsection{case 1: B-L preserving decay $n \to \bar{\nu}$:}
To begin, we consider a neutron decay scenario induced by two scalar dark matter (DM) fields, $\Phi_1$ and $\Phi_2$, corresponding to the channel $n + \Phi_1 \rightarrow \bar{\nu} + \Phi_2$. The quark-level Lagrangian governing this decay is given by,
\begin{equation}\label{uddl}
    \mathcal{L}_{udd\bar{\nu}} = \frac{1}{\Lambda_{udd \rightarrow \bar{\nu}}^4} \hspace{0.1cm} u_r d_r Q_l L_l \Phi_1^{\dagger} \Phi_2
\end{equation}
where the charges of the fields, under the gauge groups, are assigned in Table.~(\ref{tab:modelB-L}). As mentioned in the introduction, the anomalies associated with gauging $U(1)_B$ and $U(1)_L$ are assumed to be canceled~\cite{FileviezPerez:2010gw,FileviezPerez:2024fzc,Green:1984sg,Anastasopoulos:2006cz,Coriano:2007fw}. Both the dark sector fields considered here are scalars and are nearly degenerate in mass for stability reasons. 
\begin{table}[h]
    \centering
    \begin{tabular}{|c|c|c|c|c|c|}
    \hline
         & $SU(3)_c$ &  $SU(2)_W$ &  $U(1)_Y$ &  $U(1)_{B}$ &  $U(1)_{L}$  \\ \hline
         $(u_r,d_r)$& 3 & 1 & (4/3, -2/3) & 1/3 & 0\\
         $Q_l$ & 3 & 2 & 1/3 & 1/3 & 0\\
         $L_l$ & 1 & 2 & 1 & 0 & 1 \\
       $\Phi_1$  & 1 & 1 & 0 & 1 & 0 \\
       $\Phi_2$  & 1 & 1 & 0 & 0 & -1 \\
       \hline
    \end{tabular}
    \caption{Charge assignment of the SM and dark matter fields.}
    \label{tab:modelB-L}
\end{table}

With the iso-spin symmetry, the left-handed down quark in the doublet can be replaced with a left-handed up quark and the anti-neutrino can be replaced by a positron. Thus, the proton decay channel, $p + \Phi_1  \rightarrow e^+ + \Phi_2$, can also be studied in the same way as the neutron decay channel.

\begin{figure}[h]
    \centering
    \includegraphics[angle=90,width= 6 cm]{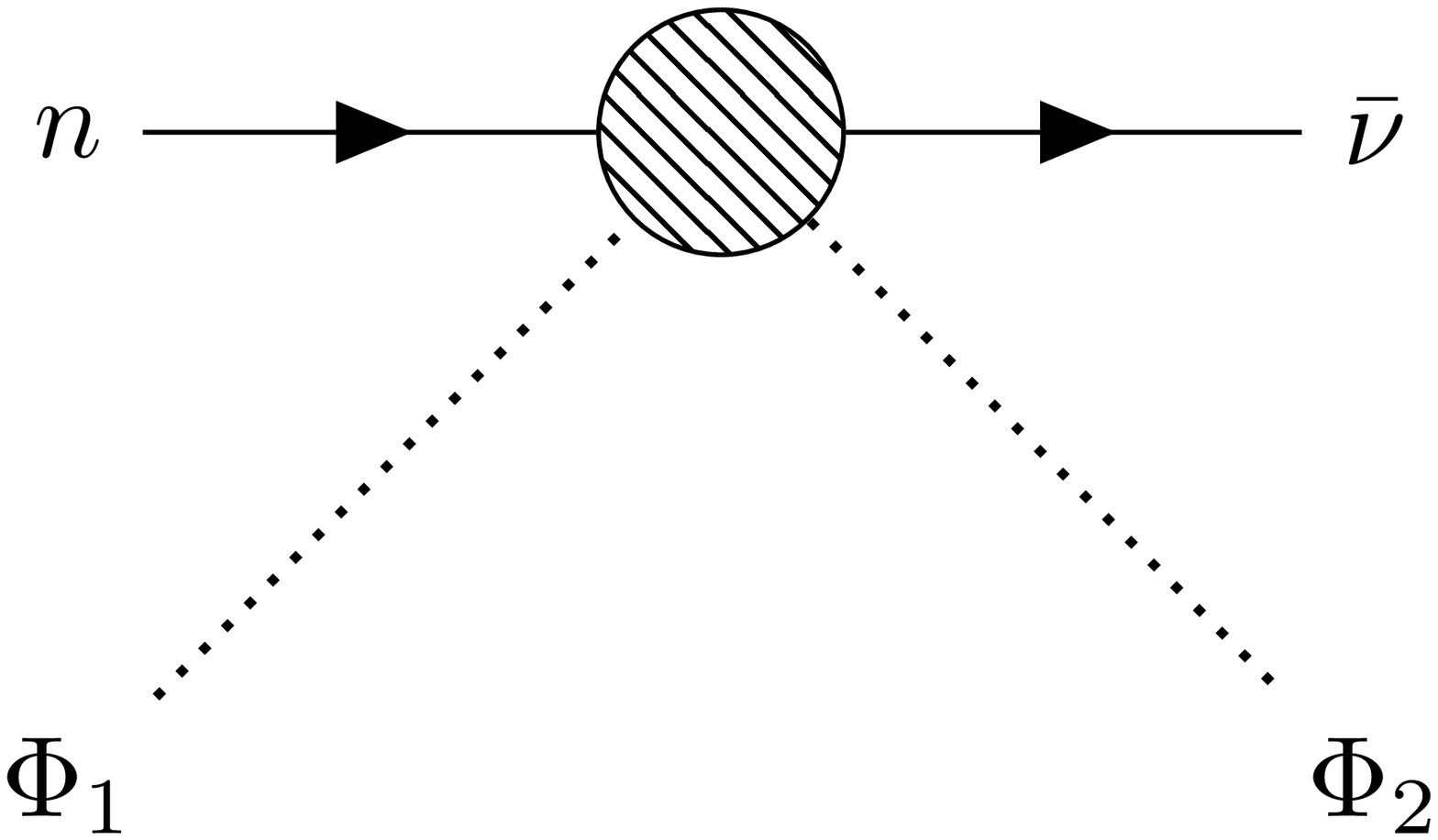}
    \caption{The decay of $n \rightarrow \bar{\nu}$}
    \label{fig 1}
\end{figure}

The parton-level operator for neutron decay given above can be matched with the hadron level operator,
\begin{equation}
    \mathcal{L}_{n\bar{\nu}} = \hspace{0.05cm} \frac{1}{\Lambda_{n \rightarrow \bar{\nu}}} \hspace{0.1cm} n \ \bar{\nu}_l \hspace{0.1cm} \Phi_1^{\dagger} \Phi_2 \ ,
    \label{eq:nbarnu}
\end{equation}
where, $\frac{1}{\Lambda_{n \rightarrow \bar{\nu}}} = \frac{\alpha}{\Lambda_{udd \rightarrow \bar{\nu}}^4} $, is the scale of new physics and $\alpha =  0.015 \hspace{0.1cm} \mathsf{GeV}^3$ gives the hadronic matrix element~\cite{Claudson:1981gh}. This operator generates the induced decay process illustrated in Fig.~(\ref{fig 1}).

The thermal cross-section for the process $n + \Phi_1 \rightarrow \bar{\nu} + \Phi_2 $ at hadron level then becomes,
\begin{equation}
    (\sigma v)_{n \rightarrow \bar{\nu}} = \frac{1}{4 \pi}
\frac{1}{\Lambda_{n \rightarrow \bar{\nu}}^2}\frac{(m_n/M_{DM})^2}{( 1- m_n/M_{DM})} \ ,
\label{eq:thermalcrossnnu}
\end{equation}
where $m_n$ is the mass of the neutron.

\subsection{case 2 : B-L violating decay $n \to \pi^0$}
\begin{figure}[h]
    \centering
    \includegraphics[angle=90,width= 6 cm]{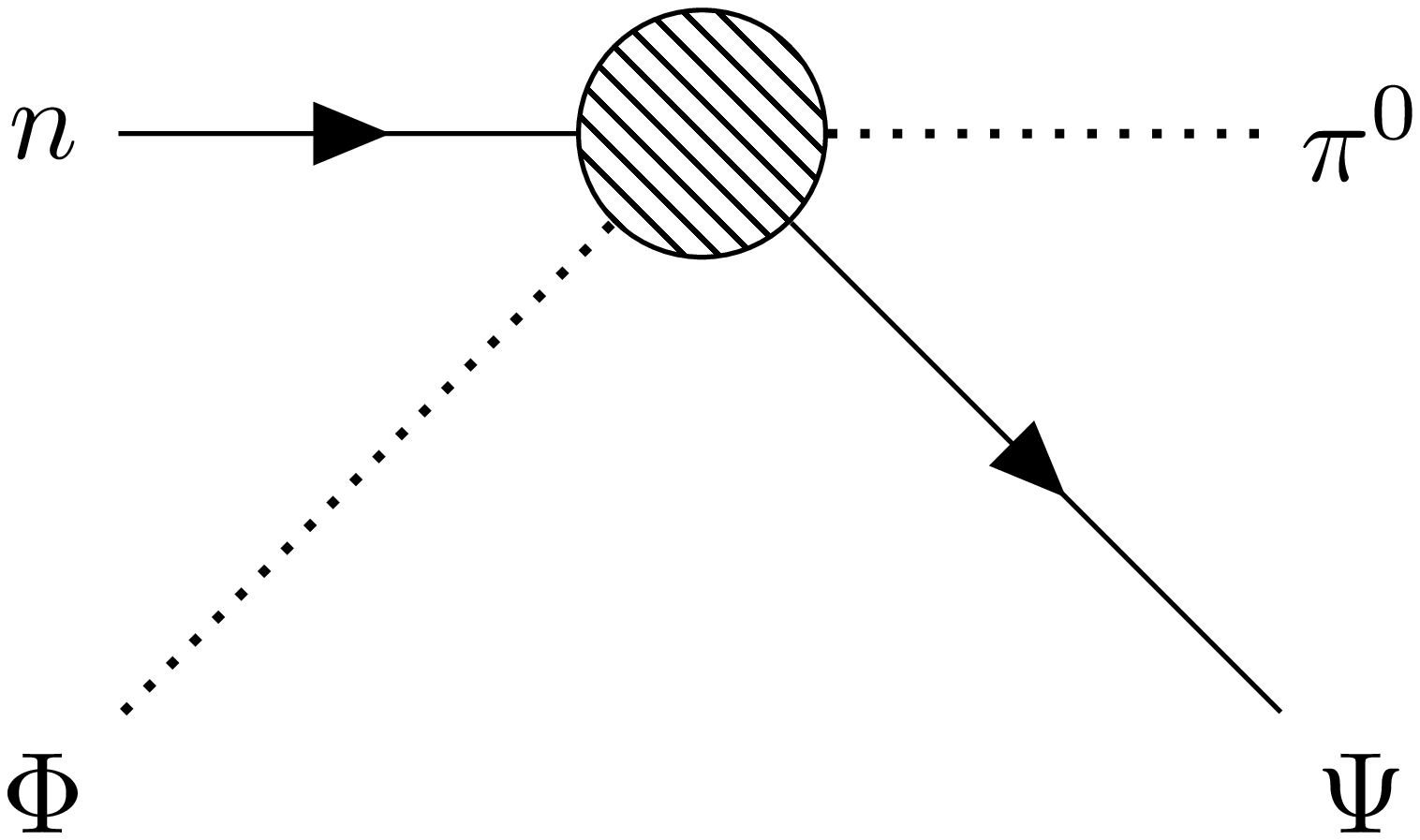}
    \caption{The decay of $n \rightarrow \pi^0$}
    \label{fig 2}
\end{figure}

An alternative scenario, in which the $B-L$ symmetry is violated, involves neutron decay into a pion within the visible sector. The corresponding Lagrangian is given by~\cite{Davoudiasl:2010am, Davoudiasl:2011fj},
\begin{equation}
    \mathcal{L}_{udd\pi} = \frac{1}{\Lambda_{udd\pi}^3} \Phi^{\dagger} \Psi d_Rd_Ru_R + h.c. \ ,
\end{equation}
where the charges of the fields are given in Table.~(\ref{tab:modelhylogenesis}). Like before, the anomalies associated with the $U(1)_B$ are assumed to be canceled by extending the fermionic sector or via axion interactions. Here, we consider a fermionic dark sector field $\Psi$ and a scalar dark sector field $\Phi$, which are nearly degenerate in mass.
\begin{table}[h]
    \centering
    \begin{tabular}{|c|c|c|c|c|}
    \hline
         & $SU(3)_c$ &  $SU(2)_W$ &  $U(1)_Y$ &  $U(1)_{B}$  \\ \hline
       $(u_r,d_r)$& 3 & 1 & (4/3, -2/3) & 1/3 \\
       $\Psi$  & 1 & 1 & 0 &  $-\frac{1}{2}$  \\
       $\Phi$  & 1 & 1 & 0 & $\frac{1}{2}$ \\
       \hline
    \end{tabular}
    \caption{Charge assignment of the dark sector fields.}
    \label{tab:modelhylogenesis}
\end{table}
On the other hand, since we are interested in low-energy experiments, the nucleon-level interaction term is given by,
\begin{equation}
    \mathcal{L}_{n\to \pi} = \frac{1}{\Lambda_{n\to\pi}} \Phi^{\dagger} \pi^0 \Psi  n\ ,
    \label{eq:ntopi}
\end{equation}
where the Feynman diagram for the induced decay is shown in Fig.~(\ref{fig 2}).
The hadronic operator is matched with the quark-level operator to get,
\begin{equation}
    \frac{1}{\Lambda_{n\to\pi}} = \frac{\alpha}{F_\pi \Lambda_{udd\pi}^3}
\end{equation}
where, $F_{\pi} = 93 \hspace{0.1cm} \mathsf{MeV}$ is the pion decay constant. And, the thermal cross-section for the process $n + \Phi \to \pi^0 + \Psi$ becomes, 
\begin{equation}
    (\sigma v)_{n \rightarrow \pi^0} = \frac{5}{16 \pi}  \frac{1}{\Lambda_{n\to\pi}^2} \frac{m_n/M_{DM}}{(1 - m_{n}/M_{DM}) } \sqrt{(1 - (m_{\pi}/m_{n})^2)} \ ,
    \label{eq:thermalcrossnpi}
\end{equation}
where $m_{\pi}$ is the masses of neutral pion. 

The induced decay width given in Eq.~(\ref{decay}) can now be computed by using Eq.~(\ref{eq:thermalcrossnnu}) and Eq.~(\ref{eq:thermalcrossnpi}) as,
\begin{eqnarray}
    \Gamma_{n\to\bar{\nu}} &=& \frac{\rho_{DM}}{M_{DM}} \hspace{0.1cm} \frac{1}{4 \pi}
\frac{1}{\Lambda_{n \rightarrow \bar{\nu}}^2}\frac{(m_n/M_{DM})^2}{( 1- m_n/M_{DM})} , \nonumber \\
 \Gamma_{n\to\pi} &=& \frac{\rho_{DM}}{M_{DM}} \hspace{0.1cm} \frac{5}{16 \pi}  \frac{1}{\Lambda_{n\to\pi}^2} \frac{m_n/M_{DM}}{(1 - m_{n}/M_{DM}) } \sqrt{(1 - (m_{\pi}/m_{n})^2)} \ .
 \label{eq:decaywidthndecay}
\end{eqnarray}


The limits on the new physics scale can be obtained by combining the experimental constraints on the nucleon decay widths listed in Table.(\ref{tab:processes}) with the above expression for the induced decay width. In particular, using the current experimental bounds from Table.(\ref{tab:processes}), the resulting constraints on the scales associated with the induced neutron decay channels ($n \rightarrow \bar{\nu}$ and $n \rightarrow \pi$), for a benchmark dark matter density of $\rho_{DM} = 0.3\ \mathsf{GeV/cm^3}$ and dark matter mass $M_{DM} = 2\ \mathsf{GeV}$, are summarized in Table.~(\ref{tab:wocor}). Since these constraints rely sensitively on both the local anti-baryonic dark matter density and the assumed scale of new physics, complementary and independent probes are therefore highly desirable.
\begin{table}[h]
    \centering
    \begin{tabular}{|c|c|c|c|}\hline
        \hspace{0.3cm} Models\hspace{0.3cm} & \hspace{0.3cm}Decay Reaction\hspace{0.3cm} & \hspace{0.3cm}$\Lambda_n$\hspace{0.3cm} & \hspace{0.3cm}$\sigma v$\hspace{0.3cm}  \\
        & & $(\mathsf{GeV})$ & $(\mathsf{GeV^{-2}})$ \\ \hline
       Case- 1  & $n \rightarrow \bar{\nu}$  & $1.03 \times 10^9$  & $3.12 \times 10^{-20}$ \\
       Case- 2  & $n \rightarrow \pi^0$  & \hspace{0.3cm}$7.31 \times 10^{10}$ \hspace{0.3cm} & \hspace{0.3cm}$1.63 \times 10^{-23}$ \hspace{0.3cm} \\ \hline
    \end{tabular}
    \caption{Limit on induced neutron decay scales. }
    \label{tab:wocor}
\end{table}


\section{Spontaneous breaking of dark baryon number and generation of di-nucleon decay}
\label{sec:SSB}


Recent theoretical developments have inspired growing interest in the spontaneous breaking of baryon and lepton number symmetries within the dark sector. This interest is motivated by a range of phenomena, including baryogenesis via scalar fields coupled to dark matter~\cite{Sakstein:2017lfm}, the generation of dark-sector CP violation~\cite{Carena:2018cjh, Carena:2019xrr}, dark phase transitions~\cite{Biermann:2022meg}, exotic states such as the bajoron~\cite{Bittar:2024nrn}, and the production of gravitational waves~\cite{Addazi:2024cqz}. Spontaneously broken baryon and lepton symmetries can lead to observable signatures, notably neutron–antineutron oscillations and di-nucleon decay processes, which are actively sought in current experimental programs.

In the framework of asymmetric dark matter models, the operators presented in Eq.~(\ref{eq:hadronindBNV}) induce, at the one-loop level, the process $nn\to \pi\pi$, which can lead to distinctive experimental signatures involving back-to-back Cherenkov radiation. This becomes particularly relevant upon incorporating the spontaneous breaking of baryon and lepton number symmetries. Such signatures are of notable interest, as they align with current experimental search strategies. Furthermore, observation of this process would provide an independent probe of the underlying new physics scale, complementing the constraints obtained from Eq.~(\ref{eq:decaywidthndecay}).

\subsection{case 1: $\Delta B= 2 =\Delta L$ process $nn\to \bar{\nu}\bar{\nu}$}

\begin{figure}[h]
    \centering
    \includegraphics[angle=90,width= 7 cm]{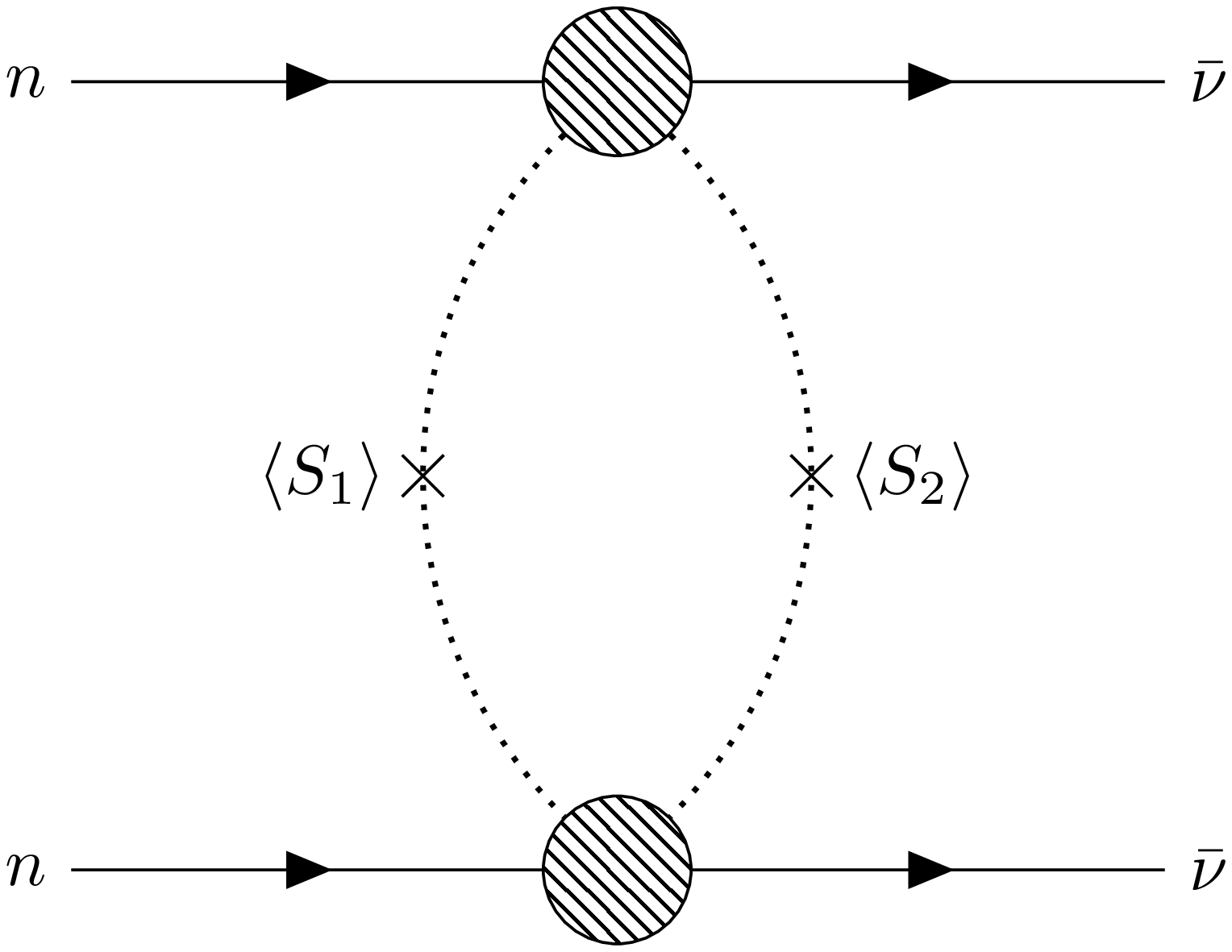}
    \caption{Di-nucleon decay with $\Delta B = 2$ and $\Delta L = 2$}
    \label{fig 3}
\end{figure}

Spontaneous breaking in the dark sector can be introduced by including two scalar fields, $S_1$ and $S_2$, charged under $U(1)_B$ and $U(1)_L$ respectively. The charges are arranged such that their interaction with the dark matter fields are given by, 
\begin{equation}
    \mathcal{L}_{int} =\frac{1}{2}\Big( \lambda_1 m_{\phi_1} S_1 \Phi_1^2 + \lambda_2 m_{\phi_2} S_2 \Phi_2^2+ \text{c.c} \Big) + m_{\phi_1}^2 \Phi_1^\dagger \Phi_1 + m_{\phi_2}^2 \Phi_2^\dagger \Phi_2 \ .
    \label{eq:s1s2}
\end{equation}
Upon spontaneously breaking the $U(1)_B$ and $U(1)_L$ symmetries through vacuum expectation values of $v_{S_1}=\langle S_1 \rangle$ and $v_{S_2}=\langle S_2 \rangle$, di-nucleon decay, $nn\to\bar{\nu}\bar{\nu}$ arise at one-loop from Eq.~(\ref{eq:nbarnu}) and Eq.~(\ref{eq:s1s2}) as illustrated in Fig.~(\ref{fig 3}). The associated Goldstone bosons, in the unitary gauge, become non-dynamical or gets "eaten away" by the gauge fields of $U(1)_B$ and $U(1)_L$. Thus, there are no massless modes in the energy spectrum of the model.

The effective Lagrangian for the di-nucleon decay process at the hadron level can be written as,
\begin{equation}
   \mathcal{L}_{nn\bar{\nu}\bar{\nu}} = \frac{1}{\Lambda_{nn \rightarrow \bar{\nu} \bar{\nu}}^2}(n\bar{\nu}_l)\hspace{0.1cm}(n\bar{\nu}_l) \ ,
   \label{eq:nnnunu}
\end{equation}
where $\Lambda_{nn \rightarrow \bar{\nu} \bar{\nu}}$ denotes the characteristic scale of the new physics responsible for the baryon number violating process. The one-loop process shown in Fig.~(\ref{fig 3}) generates the operator in Eq.~(\ref{eq:nnnunu}) from the induced nucleon decay operator in Eq.~(\ref{eq:nbarnu}). The loop integral is dominated by the infrared region, and upon integrating over the loop momentum, the matching relation takes the form,
\begin{eqnarray}
    \frac{1}{\Lambda_{nn \rightarrow \bar{\nu}\bar{\nu}}^2} &=& \frac{1}{16 \pi^2} \frac{1}{\Lambda_{n \rightarrow \bar{\nu}}^2} \frac{1}{6}\prod_{i=1,2}\Big(\frac{\lambda_i v_{S_i} m_{\phi_i}}{M_{DM_i}^2}\Big) \sim \frac{1}{16 \pi^2} \frac{1}{\Lambda_{n \rightarrow \bar{\nu}}^2} \frac{1}{6}\Big(\frac{\lambda_1 v_{S_1}\lambda_2 v_{S_2}}{M_{DM}^2}\Big)\Big|_{m_{\phi_i} \gg \lambda_i v_{S_i}} \nonumber \\
    &=& \frac{1}{16 \pi^2} \frac{1}{\Lambda_{n \rightarrow \bar{\nu}}^2} \frac{1}{6} R_{\bar{\nu}\bar{\nu}} \ ,
    \label{eq:nnnununnu}
\end{eqnarray}
where $M_{DM_i}^2 = \lambda_i v_{S_i} m_{\phi_i} + m_{\phi_i}^2$ for $i=1,2$, corresponding to $\Phi_1$ and $\Phi_2$. In the simplifying limit $m_{\phi_1}\simeq m_{\phi_2}$, we identify $M_{DM_1} = M_{DM_2} \equiv M_{DM}$. The dimensionless quantity, 
\begin{equation}
    R_{\bar{\nu}\bar{\nu}} = \frac{\lambda_1 v_{S_1}\lambda_2 v_{S_2}}{M_{DM}^2} \ll 1 \nonumber \ ,
\end{equation}
encapsulates the suppression associated with the spontaneous breaking of $U(1)_B \times U(1)_L$ and will be treated as a free parameter in the subsequent analysis.


From the effective Lagrangian, the decay width corresponding to the di-nucleon decay channel $nn \to \bar{\nu}\bar{\nu}$, takes the form,
\begin{equation}
\Gamma_{nn \to \bar{\nu}\bar{\nu}} \sim \frac{m_n^5}{\Lambda_{nn \to \bar{\nu}\bar{\nu}}^4} \ ,
\label{gammannnunu}
\end{equation}
where $m_n$ is the neutron mass. And the present experimental lower limit on the di-nucleon lifetime, quoted in Table~(\ref{tab:processes}), translates into the bound,
\begin{equation}
\frac{1}{\Lambda_{nn \to \bar{\nu}\bar{\nu}}^2} < 1.43 \times 10^{-31}\ \mathsf{GeV}^{-2}.
\end{equation}
In addition, by combining this result with the loop-induced matching relation discussed earlier, one can extract an independent constraint on the induced decay operator. For a benchmark choice of dark matter mass $M_{DM}=2\ \mathsf{GeV}$ and loop suppression factor $R_{\bar{\nu}\bar{\nu}} \sim 10^{-7}$, the corresponding lower bound on the effective scale can be estimated as,
\begin{equation}\label{eq:lambdannu}
\Lambda_{n \to \bar{\nu}} > 4.2 \times 10^{7}\ \mathsf{GeV}.
\end{equation}
This demonstrates that di-nucleon searches, despite being loop-suppressed, can still provide a meaningful and complementary probe of the underlying baryon-number-violating dynamics.



\subsection{case 2: $\Delta B=2$ process $nn\to \pi^0 \pi^0$}
\begin{figure}[h]
    \centering
    \includegraphics[angle=90,width= 7 cm]{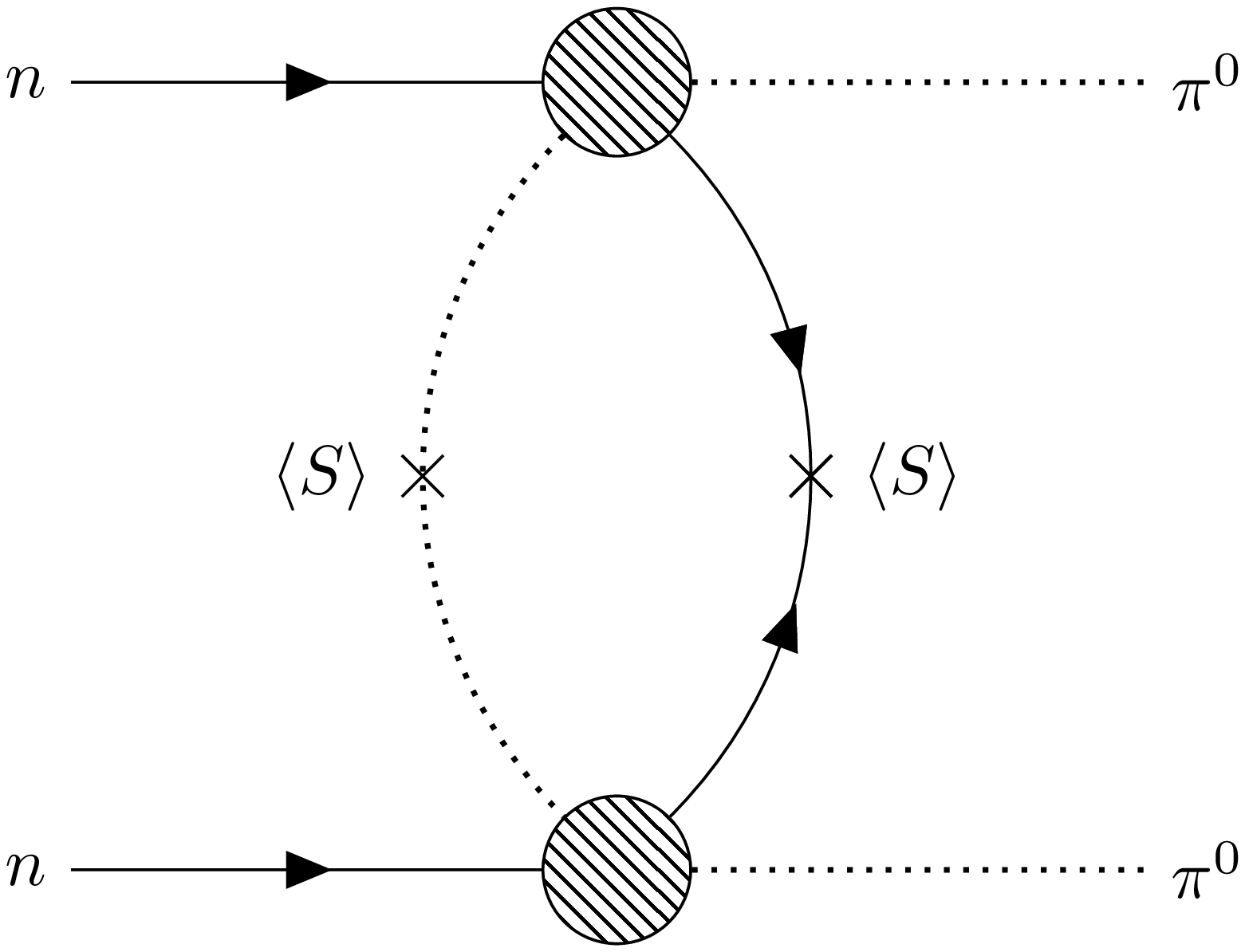}
    \caption{Di-nucleon decay with $\Delta B = 2$}
    \label{fig 4}
\end{figure}

We now consider a Standard Model singlet dark sector scalar field $S$ carrying non-zero baryon number with interactions described by,
\begin{equation}
    \mathcal{L}_{int} = \Big(\lambda_1 m_{\phi }S \Phi^2 + \lambda_2 S^\dagger \Psi^T \Psi+ c.c \Big) + m_{\phi}^2 \Phi^\dagger \Phi + m_{\psi} \Bar{\Psi} \Psi \ .
    \label{eq:sdelta}
\end{equation}
Analogous to the previous case, di-nucleon decay $nn\to\pi^0 \pi^0$ arises at one-loop from the neutron decay operators in Eq.~(\ref{eq:ntopi}), as illustrated in Fig.~(\ref{fig 4}) after the spontaneous breaking of $U(1)_B$ symmetry via the vacuum expectation values $v_{S}=\langle S \rangle$. In the unitary gauge, the gauge boson associated with the $U(1)_B$ symmetry "eats-up" the Goldstone boson to become heavy.

The hadronic effective Lagrangian governing the di-neutron decay channel $nn \to \pi^0 \pi^0$ can be expressed in terms of the baryon number violating four-fermion operator,
\begin{equation}
   \mathcal{L}_{nn\pi\pi} = \frac{1}{\Lambda_{nn \rightarrow \pi\pi}}(n n \pi^0 \pi^0) \ ,
   \label{eq:nnpipi}
\end{equation}
where $\Lambda_{nn \to \pi\pi}$ denotes the effective new-physics scale associated with the process.
Matching the effective operator in Eq.~(\ref{eq:nnpipi}) with the one-loop process in Fig.~(\ref{fig 4}) we obtain the relation,
\begin{eqnarray}\label{eq:nnpipinpi}
    \frac{1}{\Lambda_{nn \rightarrow \pi\pi}} &=& \frac{1}{16 \pi^2} \frac{1}{\Lambda_{n\to \pi}^2} \frac{1}{3}  \Big(\frac{\lambda_1\lambda_2 v_S^2 m_{\phi}}{M_{DM_1}M_{DM_2}}\Big) \nonumber \\
    &=& \frac{1}{16 \pi^2} \frac{M_{DM}}{\Lambda_{n\to \pi}^2} \frac{1}{3} R_{\pi\pi} \ ,
\end{eqnarray}
where $M_{DM_1}^2 = \lambda_1 v_S m_\phi + m_\phi^2$ and $M_{DM_2} = \lambda_2 v_S + m_\psi$ are the masses of the dark matter candidates. In the second line of the above equation, we have assumed $m_\phi \simeq m_{\psi} $ which allows us to identify $M_{DM_1}=M_{DM_2}=M_{DM}$. Furthermore, the dimensionless parameter,
\begin{equation}
    R_{\pi\pi}= \frac{\lambda_1\lambda_2 v_S^2}{M_{DM}^2} \ll 1 \nonumber \ ,
\end{equation}
like before, incorporates the suppression associated with the spontaneous breaking of the baryon symmetry.

The decay width corresponding to the di-nucleon decay channel $nn \to \pi^0\pi^0$ becomes,
\begin{equation} 
    \Gamma_{nn\pi\pi} \sim \frac{m_n^3}{\Lambda_{nn \rightarrow \pi\pi}^2} \ .
    \label{gammannpipi}
\end{equation}
Together with the experimental lower bound on the lifetime for the di-nucleon decay channel $nn \rightarrow \pi^0\pi^0$, given in Table. (\ref{tab:processes}), the constraint on the scale for this process can be computed as,
\begin{equation}
    \frac{1}{\Lambda_{nn \rightarrow \pi\pi}^2} < 6.11 \times 10^{-63} \hspace{0.1cm} \mathsf{GeV}^{-2} \ .
\end{equation}
This bound, although highly suppressed, has important implications for scenarios where baryon number violating processes are mediated through the dark sector. In particular, assuming a benchmark dark matter mass $M_{DM} = 2\ \mathsf{GeV}$ and $R_{\pi\pi} \sim 10^{-5}$ the independent lower bound on the scale of the induced neutron–pion decay operator, $\Lambda_{n \rightarrow \bar{\nu}}$ becomes,
\begin{equation}\label{eq:lambdanpi}
    \Lambda_{n \to \pi} >  1.8 \times 10^{9} \hspace{0.1cm}\mathsf{GeV} \ .
\end{equation}


The spontaneous breaking of baryon and/or lepton number in the dark sector thus gives rise to an independent constraint on the scale of the induced nucleon decay operators. In particular, the bounds derived from the di-nucleon decay channels serve as complementary probes to those obtained from single nucleon decay processes. Correlating the limits in Eq.(\ref{eq:lambdannu}) and Eq.(\ref{eq:lambdanpi}) with the decay width for induced nucleon decay in Eq.(\ref{eq:decaywidthndecay}) provides a nontrivial cross-check on the parameter space of the model. Such a correlation not only constrains the effective scale of baryon number violation, but also offers an additional handle for restricting the local density ($\rho_{DM}$), masses ($M_{DM}$) and the strength of the spontaneous symmetry breaking ($\lambda_i$) of the dark sector fields.


\section{Correlating induced nucleon decay with di-nucleon decay}
\label{sec:results}
In the previous section, we derived the di-nucleon processes generated at one-loop from the induced nucleon decay operators. Furthermore, by employing the experimental constraints on the di-nucleon decay widths in Eq.~(\ref{gammannnunu}) and Eq.~(\ref{gammannpipi}), we extracted the corresponding limits on the new physics scales for the nucleon decay operators, as shown in Eq.~(\ref{eq:lambdannu}) and Eq.~(\ref{eq:lambdanpi}). 

This implies that while the nucleon decay constraints are sensitive to local density, the di-nucleon processes provide a model independent benchmark for the effective scales of baryon number violation. Consequently, the interplay between these two sets of observables enables us to translate the bounds from di-nucleon processes into nontrivial constraints on the local dark matter abundance relevant for laboratory experiments. In the following, we quantify this relation explicitly and demonstrate how the di-nucleon bounds sharpen the consistency requirements on induced nucleon decay signatures.


For $n \rightarrow \bar{\nu} + \text{invisible}$ decay, the thermal cross-section in Eq.~(\ref{eq:thermalcrossnnu}) can be re-expressed using the bound on the new physics scale $\Lambda_{n\to\bar{\nu}}$ obtained in Eq.(\ref{eq:lambdannu}). Substituting this constraint, we get,
\begin{eqnarray}\label{thcronnu}
    (\sigma v)_{n \rightarrow \bar{\nu}} &=& \frac{1}{4 \pi}
\frac{1}{\Lambda_{n \rightarrow \bar{\nu}}^2}\frac{(m_n/M_{DM})^2}{( 1- m_n/M_{DM})} \nonumber \\
&=& 1.8 \times 10^{-17} \ \mathsf{GeV}^{-2} \ ,
\end{eqnarray}
where we have used $M_{DM}=2\ \mathsf{GeV}$. Substituting this into Eq.~(\ref{eq:decaywidthndecay}), the decay width of the induced nucleon process becomes,
\begin{eqnarray}
 \Gamma_{n \rightarrow \bar{\nu}} &=&  \frac{\rho_{DM}}{M_{DM}} \cdot 1.8 \times 10^{-17} \ \text{GeV}^{-2} \ .
 \end{eqnarray}
Comparing with the experimental upper bound on the decay width for $n \rightarrow \bar{\nu}+  \text{invisible}$, given in Table.~(\ref{tab:processes}), allows us to directly infer the local terrestrial density of dark matter. For a dark matter of mass $M_{DM}=2\ \mathsf{GeV}$, we find, $ \rho_{DM} =  0.5 \times 10^{-3} \mathsf{GeV \cdot cm^{-3}}$, which lies well below the canonical astrophysical estimate $\rho_{DM} = 0.3 \ \mathsf{GeV \cdot cm^{-3}}$. 

\begin{figure}[h]
    \centering
    \includegraphics[width= 12cm]{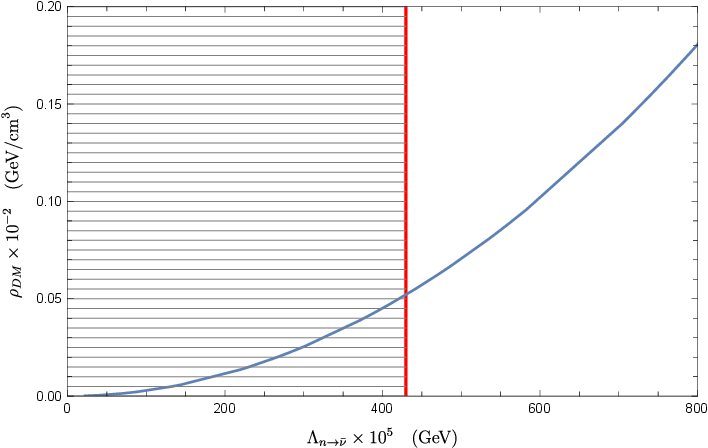}
    \caption{The plot shows the local terrestrial dark matter density $\rho_{DM}$ vs the new physics scale $\Lambda_{n\rightarrow \nu}$ with the red vertical line indicating the limit from the $nn \rightarrow \bar{\nu}\bar{\nu}$ process. The blue curve represents the parameter space that satisfies the current limit for $n \to \bar{\nu} + \text{invisible}$, assuming a dark matter mass $M_{DM}=2\ \mathsf{GeV}$~\cite{Huang:2013xfa} and $R_{\bar{\nu}\bar{\nu}}\sim 10^{-7}$. The parameter space above the curve is disallowed by $n \to \bar{\nu} + \text{invisible}$, while the shaded region is disallowed by $nn \rightarrow \bar{\nu}\bar{\nu}$.}
    \label{fig:plot1}
\end{figure}

Alternatively, instead of fixing the scale $\Lambda_{n \rightarrow \bar{\nu}}$, one can treat it as a variable and determine how the terrestrial dark matter density $\rho_{DM}$ varies as a function of $\Lambda_{n \rightarrow \bar{\nu}}$. The resulting dependence is shown in Fig.(\ref{fig:plot1}). The red vertical line denotes the lower bound on $\Lambda_{n \to \bar{\nu}}$ obtained previously in Eq.~(\ref{eq:lambdannu}). The region lying above the curve corresponds to parameter space excluded by existing nucleon decay searches, since it would predict a decay rate in excess of the experimental bounds. In addition, the shaded band indicates the portion of parameter space already disfavored by the complementary constraints arising from the di-nucleon decay channel $nn \rightarrow \bar{\nu}\bar{\nu}$.


For the case of baryon number violating decay via $n \rightarrow \pi^0$, the thermal cross-section can be computed using Eq.~(\ref{eq:thermalcrossnpi}) together with the bound on the new physics sale Eq.~(\ref{eq:lambdanpi}) as,
\begin{eqnarray}
(\sigma v)_{n \rightarrow \pi^0} &=& \frac{5}{16 \pi} \frac{1}{\Lambda_{n\to\pi}^2} \frac{m_n/M_{DM}}{(1 - m_{n}/M_{DM}) } \sqrt{1 - \left(\frac{m_{\pi}}{m_{n}}\right)^2} \nonumber \\
&=& 2.5 \times 10^{-20}\hspace{0.1cm} \mathsf{GeV}^{-2} \ .
\end{eqnarray}

Substituting this into the second expression in Eq.~(\ref{eq:decaywidthndecay}), the induced nucleon decay width becomes,
\begin{eqnarray}
\Gamma_{n \rightarrow \pi} &=& \frac{\rho_{DM}}{M_{DM}} \cdot 2.5 \times 10^{-20} \hspace{0.1cm} \mathsf{GeV}^{-2} \ .
\end{eqnarray}

Using the experimental upper limit on the partial width for $n \rightarrow \pi^0$, listed in Table.~(\ref{tab:processes}), the corresponding local dark matter density can be inferred to be $\rho_{DM} = 0.2 \times 10^{-3}\ \mathsf{GeV/cm^3}$ for a dark matter candidate with mass $2\ \mathsf{GeV}$.

As before, treating $\Lambda_{n \to \pi}$ as a variable, Fig. (\ref{fig:plot1}) illustrates the terrestrial $\rho_{DM}$ as a function of the induced nucleon decay scale. The vertical red line denotes the lower bound on $\Lambda_{n \to \pi}$ obtained in Eq.(\ref{eq:lambdanpi}). Thus, like in the previous case, the overlap between the nucleon-decay and di-nucleon constraints provides a stringent combined limit on baryon number violating dark sector interactions.

\begin{figure}[h]
    \centering
    \includegraphics[width=12cm]{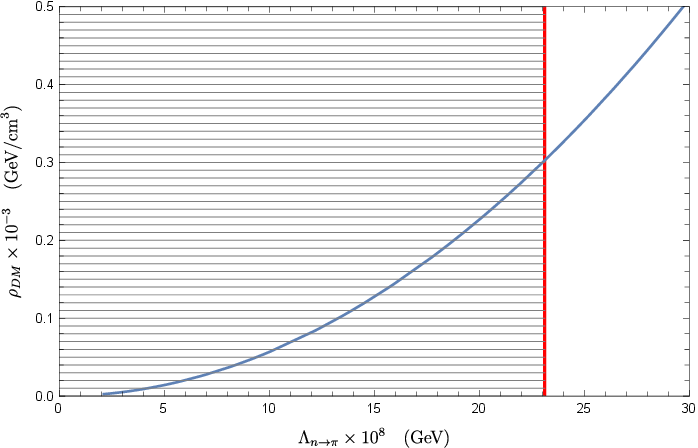}
    \caption{The plot shows the local dark matter density $\rho_{DM}$ vs the $\Lambda_{n\rightarrow \pi}$, with the red vertical line indicating the limit from the $nn \rightarrow \pi^0\pi^0$ process. The blue curve represents the parameter space that satisfies the current limit for $n \to \pi^0 + \text{invisible}$, assuming a dark matter mass $M_{DM}=2\ \mathsf{GeV}$  \cite{Davoudiasl:2010am,Davoudiasl:2011fj} and $R_{\pi\pi}\sim 10^{-5}$. The region above the blue curve is disallowed by $n \to \pi^0 + \text{invisible}$, while the shaded region is disallowed by $nn \rightarrow \pi^0\pi^0$.}
    \label{fig:plot2}
\end{figure}

Therefore, for the $B - L$ preserving process $n \rightarrow \bar{\nu}$ and the $B-L$ violating process $n\rightarrow \pi^0$, the corresponding anti-baryonic dark matter density are summarized in Table.~(\ref{tab:rhoconstraints}), assuming a benchmark dark matter mass of $M_{DM}=2\ \mathsf{GeV}$. 
\begin{table}[h]
\centering
     \begin{tabular}{|c|c|c|c|c|}\hline
      \hspace{0.2cm}Models\hspace{0.2cm} & \hspace{0.2cm} Decay Reaction \hspace{0.2cm}& \hspace{0.2cm}$\Lambda_n$ \hspace{0.2cm}& \hspace{0.2cm}$\sigma v$ \hspace{0.2cm}& \hspace{0.2cm}$\rho_{DM}$ \hspace{0.2cm}\\
       &  &$\hspace{0.2cm}(\mathsf{GeV})\hspace{0.2cm}$& $\hspace{0.2cm}(\mathsf{GeV}^{-2})\hspace{0.2cm}$ & $\hspace{0.2cm}(\mathsf{GeV \cdot cm^{-3}})\hspace{0.2cm}$ \\ \hline
      Case-1 & $n \rightarrow \bar{\nu}$ & \hspace{0.2cm} $4.3 \times 10^{7}$ \hspace{0.2cm} & $\hspace{0.2cm} 1.8 \times 10^{-17} \hspace{0.2cm}$ & $\hspace{0.2cm} 0.5 \times 10^{-3} \hspace{0.2cm}$ \\ 
      Case-2 & $n \rightarrow \pi^0$ & \hspace{0.2cm} $2.2 \times 10^{9}$ \hspace{0.2cm}  & $\hspace{0.2cm}2.5 \times 10^{-20}\hspace{0.2cm}$ & $\hspace{0.2cm}0.3 \times 10^{-3}\hspace{0.2cm}$ \\ \hline
 \end{tabular}
   \caption{Benchmark points on the constraint on local dark matter density ($\rho_{DM}$) upon simultaneously satisfying the induced nucleon and di-nucleons decays as shown in Fig.~(\ref{fig:plot1}) and Fig.~(\ref{fig:plot2}) for dark matter mass $M_{DM}=2\ \mathsf{GeV}$~\cite{Davoudiasl:2010am,Davoudiasl:2011fj,Huang:2013xfa}, assuming $R_{\bar{\nu}\bar{\nu}}\sim 10^{-7}$ and $R_{\pi\pi}\sim 10^{-5}$ respectively. }
 \label{tab:rhoconstraints}
 \end{table}
 It is thus evident that the contribution from these anti-baryonic dark matter candidates constitutes only a small fraction of the total dark matter flux at Earth. While the nucleon decay constraints yield valuable insights into the scale of baryon-number violating operators, the parameter space consistent with terrestrial nucleon decay bounds leads to suppressed dark matter densities that cannot account for the entire local abundance.
 
\begin{figure}[h]
    \centering
    \includegraphics[width=12cm]{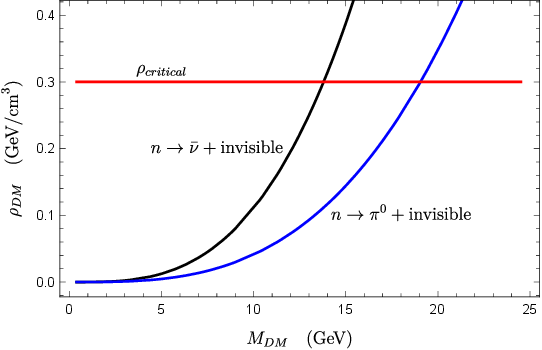}
    \caption{The plot shows the terrestrial dark matter density $\rho_{DM}$ as a function of the dark matter mass $M_{DM}$. The black (blue) curves represents the regions of parameter space that simultaneously satisfies the limits from $n\to \bar{\nu} + \text{invisible}$ and $nn \rightarrow \bar{\nu} \bar{\nu}$ ($n\to \pi^0 + \text{invisible}$ and $nn \rightarrow \pi\pi$). The red horizontal line denotes the maximum local dark matter density on Earth.}
    \label{fig:rhovsMDM}
\end{figure}

On the other hand, if we assume that the anti-baryon number charged dark matter constitutes the entirety of the local terrestrial dark matter flux ($\rho_{DM}=0.3\  \mathsf{GeV/cm^3}$), then the combined constraints from the processes $n\to \bar{\nu} + \text{invisible}$ and $nn\to \nu \nu$ bounds the dark matter mass to $\leq 14 \ \mathsf{GeV}$ as shown by the intersection of the black curve in the Fig.~(\ref{fig:rhovsMDM}) with the critical density $\rho_{critical}$. Meanwhile, as indicated by the blue curve in the same figure, the constraint from $n\to \pi^0 + \text{invisible}$ and $nn\to \pi^0 \pi^0$ restricts the dark matter mass to $\leq 20  \ \mathsf{GeV}$. These results are summarized in the Table.~(\ref{tab:MDMconstraints}).

\begin{table}[h]
\centering
     \begin{tabular}{|c|c|c|c|}\hline
      \hspace{0.2cm}Models\hspace{0.2cm} & \hspace{0.2cm} Decay Reaction \hspace{0.2cm}& \hspace{0.2cm}$\sigma v$ \hspace{0.2cm}& \hspace{0.2cm}$M_{DM}$ \hspace{0.2cm}\\
       &  & $\hspace{0.2cm}(\mathsf{GeV}^{-2})\hspace{0.2cm}$ & $\hspace{0.2cm}(\mathsf{GeV})\hspace{0.2cm}$ \\ \hline
      Case-1 & $n \rightarrow \bar{\nu}$ & $\hspace{0.2cm} 2 \times 10^{-19} \hspace{0.2cm}$ & $\hspace{0.2cm}  15 \hspace{0.2cm}$ \\ 
      Case-2 & $n \rightarrow \pi^0$ & $\hspace{0.2cm}61.4 \times 10^{-22}\hspace{0.2cm}$ & $\hspace{0.2cm} 20 \hspace{0.2cm}$ \\ \hline
 \end{tabular}
   \caption{The upper limit on mass of the anti-baryonic dark matter ($M_{DM}$) upon satisfying total dark matter flux at Earth as shown in Fig.~(\ref{fig:rhovsMDM}).}
 \label{tab:MDMconstraints}
 \end{table}

Moreover, the dependence of the local density of dark matter on $R_{\bar{\nu}\bar{\nu}}$ and $R_{\pi \pi}$ can be seen in Table.(\ref{tab:Rheirarchies}). It shows the lower limit on the local densities for dark matter of mass $M_{DM}=2 \mathrm{GeV}$. 
\begin{table}[h]
\centering
     \begin{tabular}{|c|c|c|}\hline
      \hspace{0.2cm}$R_{\bar{\nu}\bar{\nu}}$\hspace{0.2cm} & \hspace{0.2cm} $\Lambda_{n}$ \hspace{0.2cm} & \hspace{0.2cm}$\rho_{DM}$ \hspace{0.2cm}\\
      & $\hspace{0.2cm}(\mathsf{GeV})\hspace{0.2cm}$ & $\hspace{0.2cm}(\mathsf{GeV/cm^3})\hspace{0.2cm}$ \\ \hline
       $10^{-5}$ & $\hspace{0.2cm} 8.5 \times 10^{8} \hspace{0.2cm}$ & $\hspace{0.2cm}  0.2 \hspace{0.2cm}$ \\ 
       $10^{-6}$ & $\hspace{0.2cm} 8.5 \times 10^{7} \hspace{0.2cm}$ & $\hspace{0.2cm}  0.002 \hspace{0.2cm}$ \\ 
       $10^{-7}$ & $\hspace{0.2cm} 8.5 \times 10^{6} \hspace{0.2cm}$ & $\hspace{0.2cm}  0.2 \times 10^{-4} \hspace{0.2cm}$ \\ \hline
 \end{tabular}
 \qquad
     \begin{tabular}{|c|c|c|}\hline
      \hspace{0.2cm}$R_{\pi\pi}$\hspace{0.2cm} & \hspace{0.2cm} $\Lambda_{n}$ \hspace{0.2cm} & \hspace{0.2cm}$\rho_{DM}$ \hspace{0.2cm}\\
       & $\hspace{0.2cm}(\mathsf{GeV})\hspace{0.2cm}$ & $\hspace{0.2cm}(\mathsf{GeV/cm^3})\hspace{0.2cm}$ \\ \hline
       $3\times 10^{-4}$ & $\hspace{0.2cm} 7 \times 10^{10} \hspace{0.2cm}$ & $\hspace{0.2cm}  0.27 \hspace{0.2cm}$ \\ 
       $3\times 10^{-5}$ & $\hspace{0.2cm} 7 \times 10^{9} \hspace{0.2cm}$ & $\hspace{0.2cm}  0.3\times 10^{-2} \hspace{0.2cm}$ \\ 
       $3\times 10^{-6}$ & $\hspace{0.2cm} 7 \times 10^{8} \hspace{0.2cm}$ & $\hspace{0.2cm}  0.3\times 10^{-4} \hspace{0.2cm}$ \\ \hline
 \end{tabular}
 
   \caption{The lower limit on local density of the anti-baryonic dark matter ($M_{DM} =2$) for different $R_{\bar{\nu}\bar{\nu}}$ and $R_{\pi \pi}$.}
 \label{tab:Rheirarchies}
 \end{table}

\section{Summary}
\label{sec:summary}
Baryon number violation is a key ingredient in generating the observed baryon asymmetry in the universe. Since the Standard Model conserves baryon number at the perturbative level, observing these rare decay processes would provide clear evidence of New Physics. On the other hand, current experimental constraints place the proton/neutron lifetime above $ 10^{34} \hspace{0.1cm} \text{years}$, setting stringent limits on models with baryon-violating processes. Complementary to this, di-nucleon decay, which involves the simultaneous decay of two bound nucleons, provides an alternative probe of baryon number violation. Processes that produce back-to-back Cherenkov radiations, such as $ pp \to \pi^+ \pi^+ $, $ nn \to \pi^0 \pi^0 $ etc, are, hence, actively searched by current experiments. The operators that generate such di-nucleon processes are interesting as their Wilson coefficients are less suppressed. In addition to visible decay channels, invisible neutron decay like $n\to \bar{\nu}+ \text{invisible}$, and $nn\to \bar{\nu}\bar{\nu}$, provide another potential signature of baryon number violation. Experimental searches for unstable nuclei in large volume detectors aim to detect such decays, which, if observed, could provide crucial insights into the nature baryon number violation.

In this article, we explore how di-nucleon decay is generated with spontaneous breaking of baryon number in the dark sector. While the spontaneous breaking of baryon number in the dark sector has been of interest in the context of dark CP violation and dark phase transitions, we observe in our paper that induced nucleon decay generated by anti-baryonic dark matter and di-nucleon decays are correlated. Our analysis demonstrates that the effective field theory (EFT) operators responsible for induced nucleon decay ($n \to \bar{\nu} + \text{invisible}, \ n \to \pi^0 + \text{invisible}$) also generate di-nucleon decay processes ($nn \to \bar{\nu}\bar{\nu}, \ nn \to \pi^0 \pi^0$) at the one-loop level. This connection provides a complementary approach to probe induced baryon number violation, which plays a crucial role in understanding the origin of baryon asymmetry and its possible link to dark matter.

Based on current experimental constraints from KamLand and Super-Kamiokande for di-nucleon decay, the analysis constrains the new physics scale of the induced nucleon decay operator in the range of $ \mathcal{O}(10^{7} \mathsf{GeV})$, depending on the specific decay mode. These limits are competitive with those derived from traditional nucleon decay searches, reinforcing the viability of di-nucleon decay as a complementary and potentially more accessible probe of baryon number violation, especially since it predicts a back-to-back Cherenkov radiation signature. 

For the $B-L$ preserving decay channel, the fraction of anti-baryonic dark matter density in the total dark matter flux at Earth is computed to be $\rho_{\text{DM}} = 0.2 \ \mathsf{GeV/cm^3}$ for dark matter with mass $M_{\text{DM}} = 2 \mathrm{GeV}$ and $R_{\bar{\nu}\bar{\nu}}=10^{-5}$. In contrast, for the $B-L$ violating decay, the dark matter density becomes $\rho_{\text{DM}} = 0.3 \times 10^{-2} \ \mathsf{GeV}/\mathsf{cm}^3$ for a similar value of $R_{\pi \pi}$. These values indicate that neutron decay channels provide an indirect probe of the local dark matter environment and can impose limits on the anti-baryonic dark matter density.  On the other hand, if we assume that this dark matter accounts for the entire flux on Earth, the upper limit on the mass of dark matter is strongly constrained to be approximately $14 \ \mathsf{GeV}$ and $20\ \mathsf{GeV}$ for the $B-L$ preserving and $B-L$ violating scenarios, respectively.

\section*{Acknowledgments}
M.T.A. acknowledges the financial support of DST through the INSPIRE Faculty grant DST/INSPIRE/04/2019/002507. A.B.K. thanks IISER Thiruvananthapuram for hosting her during this project.

\bibliographystyle{unsrt}

\bibliography{bib}

\end{document}